\documentclass[twocolumn]{aastex6}
\usepackage{amsmath,color,textcomp,url,graphicx,subfigure}

\newcommand{\kms}{\hbox{km\,s$^{-1}$}}
\newcommand{\mjup}{$M_\mathrm{Jup}$}

\slugcomment{To appear in ApJ}

\shorttitle{WISE J1355$-$8258: A Possible Young Spectral Binary}
\shortauthors{Bardalez Gagliuffi et al.}

\begin{document}

\title{An L+T Spectral Binary with Possible AB~Doradus Kinematics}

\author{Daniella C. Bardalez Gagliuffi\altaffilmark{1,3,4}, Jonathan Gagn\'e\altaffilmark{2,5}, Jacqueline K. Faherty\altaffilmark{3}, Adam J. Burgasser\altaffilmark{1}}
\affil{\altaffilmark{1} Center for Astrophysics and Space Sciences, University of California, San Diego, 9500 Gilman Dr., Mail Code 0424, La Jolla, CA~92093, USA; daniella@physics.ucsd.edu\\
\altaffilmark{2} Carnegie Institution of Washington DTM, 5241 Broad Branch Road NW, Washington, DC~20015, USA\\
\altaffilmark{3} Department of Astrophysics, American Museum of Natural History, Central Park West at 79th St., New York, NY, 10024, USA\\
\altaffilmark{4} AMNH Kalbfleisch Fellow\\
\altaffilmark{5} NASA Sagan Fellow\\}



\begin{abstract}
We present the identification of WISE~J135501.90$-$825838.9 as a spectral binary system with a slight possibility of planetary-mass components in the $130-200$\,Myr AB~Doradus moving group. Peculiarities in the near-infrared spectrum of this source suggest it to be a blended-light binary with L6.0$\pm$1.0 and T3.0$\pm$1.8 or L7.0$\pm$0.6 and T7.5$\pm$0.4 components. Its proper motion and radial velocity as a combined-light source yield a high membership probability for AB~Doradus. While the young L6+T3 case is underluminous in a color-magnitude diagram at the AB~Doradus kinematic distance, the young L7+T7.5 case could be viable. Gravity-sensitive indicators are more consistent with a field-age binary. If confirmed as a young object, member of AB~Doradus, we estimate masses of 11$\pm$1\,\mjup\ and 9$\pm$1\,\mjup\ with both component masses below the Deuterium burning mass limit. Otherwise, we find masses of  $72^{+4}_{-5}$\,\mjup\ and $61^{+6}_{-8}$\,\mjup\ for the field L6+T3 case and $70^{+2}_{-4}$\,\mjup\  and $42^{+5}_{-6}$\,\mjup\ for the field L7+T7.5 case. Our identification of WISE~J135501.90$-$825838.9 as a candidate young spectral binary introduces a new technique for detecting and characterizing planetary-mass companions to young brown dwarfs.
\end{abstract}




\section{Introduction} \label{sec:intro}

Brown dwarfs span the mass range between the hydrogen and deuterium burning mass limits (13\,\mjup~$< M <$~80\,\mjup;~\citealt{1973ApJ...180..195G,2011ApJ...727...57S,2001RvMP...73..719B}). While the inability to sustain hydrogen fusion and consequent long-term cooling differentiates brown dwarfs from stars, distinguishing low-mass brown dwarfs from giant planets is less straightforward. Signatures from different formation pathways leading to giant exoplanets or brown dwarfs (e.g. core accretion or gravitational collapse) vanish within a couple hundred million years~\citep{2007ApJ...655..541M}, while the direct detection of deuterium is inhibited by its low cosmic abundance (D/H $= (2.45\pm0.10)\times10^{-5}$;~\citealt{2015PhRvD..92l3526C}). Non-solar initial compositions could distinguish exoplanet and brown dwarf spectra, but such abundance differences have yet to be widely demonstrated~\citep{2013Sci...339.1398K}. Alternative hypotheses for the formation of both low-mass brown dwarfs and high-mass giant planets (e.g. CoRoT-Exo-3b;~\citealt{2008A&A...491..889D}) postulate formation as a consequence of planetary collisions, leading to compact objects on highly eccentric orbits~\citep{2008A&A...482..315B}.


Most of the directly-imaged giant exoplanets discovered to date have spectral types in the mid-L to early-T range, and are companions to young stars (e.g.~\citealt{2015Sci...350...64M}).  Isolated L and T dwarfs in young moving groups are useful proxies for studying giant exoplanet atmospheres. The BASS-Ultracool Survey (BASS-UC;~\citealt{2015ApJ...808L..20G}, Gagn\'{e} et al., in prep.) searches for the coolest late-L and T-type members of young moving groups using the Bayesian Analysis for Nearby Young AssociatioNs II (BANYAN~II;~\citealt{2014ApJ...783..121G,2013ApJ...762...88M}) tool to assign membership probability based on sky position and kinematics. Targets for BASS-UC are selected from a photometric crossmatch of the 2MASS~\citep{2006AJ....131.1163S} and AllWISE~\citep{2010AJ....140.1868W} catalogs with restrictions on proper motion ($\mu_{\mathrm{total}}>$30\,mas yr$^{-1}$) and customized color cuts. Sources are selected as young moving group candidates if their BANYAN~II Bayesian probability exceeds $90\%$ and their optimal position in $UVW$ velocity space is within 5\,\kms~of the mean motion of the most probable moving group. This program identified the T5.5 SDSS~J111010.01+011613.1~\citep{2004AJ....127.3553K} as a member of the AB Doradus moving group ($130-200$ Myr;~\citealt{2015ApJ...808L..20G,2015MNRAS.454..593B}), implying an estimated mass ($\sim10-12$\,\mjup) below the deuterium burning limit.

In this paper, we present the identification of WISE~J135501.90$-$825838.9 (hereafter WISE~J1355$-$8258) as a young spectral binary candidate with L and T-dwarf components from the BASS-UC survey. Originally discovered by~\citet{2016ApJS..224...36K} and identified as a possible subdwarf, the near-infrared spectrum of this source exhibits peculiarities attributable to unresolved binarity. Additionally, its sky position and kinematics yield a high likelihood of membership in AB Doradus. In Section~\ref{sec:obs} we describe imaging and spectroscopic observations of WISE J1355$-$8258. In Section~\ref{sec:SB} we examine evidence that this peculiar source is a spectral binary. In Section~\ref{sec:young} we qualify its possible membership in the AB Doradus moving group. In Section~\ref{sec:analysis}, we analyze color-magnitude diagrams and estimate physical properties of the system. We discuss the implications of these results, including future young giant exoplanet searches with the spectral binary technique in Section~\ref{sec:discussion}.

\section{Observations}\label{sec:obs}
WISE~J1355$-$8258 was first identified as a ``photometric'' L7.1 in a high proper motion survey~\citep{2016ApJ...817..112S} based on a crossmatch between AllWISE and the first pass of NEOWISER~\citep{2014ApJ...792...30M}. It was independently identified as a low-mass, young AB Doradus candidate in the BASS-UC survey.~\citet{2016ApJS..224...36K} obtained a near-infrared spectrum of this object and classified it as an sdL5$?$ based on its blue near-infrared spectral energy distribution (SED) compared to the L5 standard 2MASS~J08350622+1953050~\citep{2010ApJS..190..100K}, attributed to stronger $H_2$ collision induced absorption.

\subsection{FIRE Spectroscopy}\label{sec:obsfire}
We obtained new spectroscopic observations of WISE~J1355$-$8258 with the Folded-port InfraRed Echellete (FIRE;~\citealt{2008SPIE.7014E..0US}) instrument, mounted on the 6.5\,m Walter Baade Telescope, located at Las Campanas Observatory, La Serena, Chile. This object was observed in prism mode on UT 2016 January 22 with the $0\farcs6$ slit, and in echelle mode on UT 2017 June 11 with the $0\farcs6$ slit, sampling wavelengths $0.82-2.51\,\mu$m at respective resolutions of $R\sim 300-500$ and $R\sim$6000. The weather was clear on both nights with a seeing of $0\farcs3$ and $0\farcs6$, respectively. For the prism spectrum, eight exposures of 60\,s each were taken at an airmass of 1.76, immediately followed by six 1\,s exposures of the A0-type star HD~149818 at an airmass of 1.81. The A0 standard was moved slightly off the slit to avoid saturation, leading to wavelength-dependent slit losses that impeded the calibration of the relative spectral response. HeNeAr lamp exposures were obtained for wavelength calibration. For the echelle spectrum, four exposures of 400\,s each were taken at an airmass of 1.79--1.82, immediately followed by six 1\,s exposures of the A0-type star HD~113512 at an airmass of 1.71. A single 10\,s Th-Ar wavelength calibration exposure was obtained immediately after the science and A0-type acquisitions without moving the telescope. Eleven high-flux 3\,s internal flat fields were obtained at the beginning of the night and eleven low-flux 20\,s internal flat fields were obtained at the end of the night.

All data were reduced with the Interactive Data Language (IDL) Firehose v2.0 package\footnote{Available at~\url{https://github.com/jgagneastro/FireHose_v2/tree/v2.0} This package makes use of the telluric correction routines contained in SpeXtool~(\citealt{2003PASP..115..389V,2004PASP..116..362C}), and the original FireHose pipeline.}~(\citealt{2009PASP..121.1409B, zenodofirehose}) following standard procedures (See~\citealt{2015ApJS..219...33G}).  The de-centering of the A0 standard in prism mode produced wavelength-dependent slopes in the final reduced spectrum, particularly severe in the $K_s$-band. To correct for the reddened slope, we used photometry from our FourStar observations described in the next subsection.


\subsection{FourStar Imaging}

We supplemented the FIRE observations with near-infrared imaging from the FourStar Infrared Camera~\citep{2013PASP..125..654P}, also mounted on the Magellan Baade Telescope, on UT 2017 June 12 under photometric conditions with $0\farcs6$ seeing. Thirty-three 20\,s exposures were obtained following a Poisson dither pattern with 11 random positions for each of the $J$, $H$ and $K_S$ filters at airmasses in the range $1.80-1.85$. These filters cover wavelengths of $1.0-2.5\,\mu$m with a plate scale of $0\farcs159$/pixel. 

The data were reduced using the FSRED FourStar data reduction package\footnote{The package can be found at~\url{http://instrumentation.obs.carnegiescience.edu/FourStar/SOFTWARE/reduction.html}.}, using the default set of calibrations. FSRED performs flat-fielding to correct for pixel response variation, a sky subtraction from the data with stars masked, and an initial astrometric solution. Most of these steps rely on the AstrOmatic package\footnote{\url{https://www.astromatic.net/}}.

A more accurate astrometric solution was obtained for the first frame of each band with \url{astrometry.net} based on the USNO-B catalog index files, and then new index files were built for that exposure. All subsequent exposures were then anchored on the first one using \url{astrometry.net} to take full advantage of the $0\farcs159$ pixel size of the FourStar detector. The individual frames were combined into a mosaic using the IDL routine \texttt{hastrom}\footnote{From the NASA GSFC library: \url{https://idlastro.gsfc.nasa.gov/ftp/pro/astrom/hastrom.pro}}.

Aperture photometry was used to measure the $J$, $H$ and $K_{S}$-band magnitudes of WISE~J1355$-$8258 with the FourStar imaging data. We measured the relative photometry of WISE~J1355$-$8258 with respect to nearby bright sources below saturation ($J < 14$) and AAA photometric quality within the frame field of view.  Sources for aperture photometry were selected by crossmatching our FourStar images per filter with 2MASS within a radius $<3"$. The average sky flux was calculated in a 14\arcsec--28\arcsec\ aperture annulus around the target, and the sky-subtracted flux was calculated within a radius of 4\farcs6 of the target. This was repeated on roughly 8--15 reference stars (depending on the filter) with high-quality 2MASS photometry in the same FourStar detector, and a zero-point (ZP) was determined for each reference star with $m = 2.5\log_{10}{\rm{countrate}} + \rm{ZP}$, where $m$ is the 2MASS measurement and $ZP$ is the magnitude for 1 count/second. The average zero-point of reference stars was taken as that of the data within each band, and was used to transform the WISE~J1355$-$8258 fluxes into magnitudes. The resulting magnitudes are reported in Table~\ref{tab:binprops}. 




\section{A Candidate L dwarf/T dwarf Binary}\label{sec:SB}

\begin{figure*}[p]
\centering
\gridline{\fig{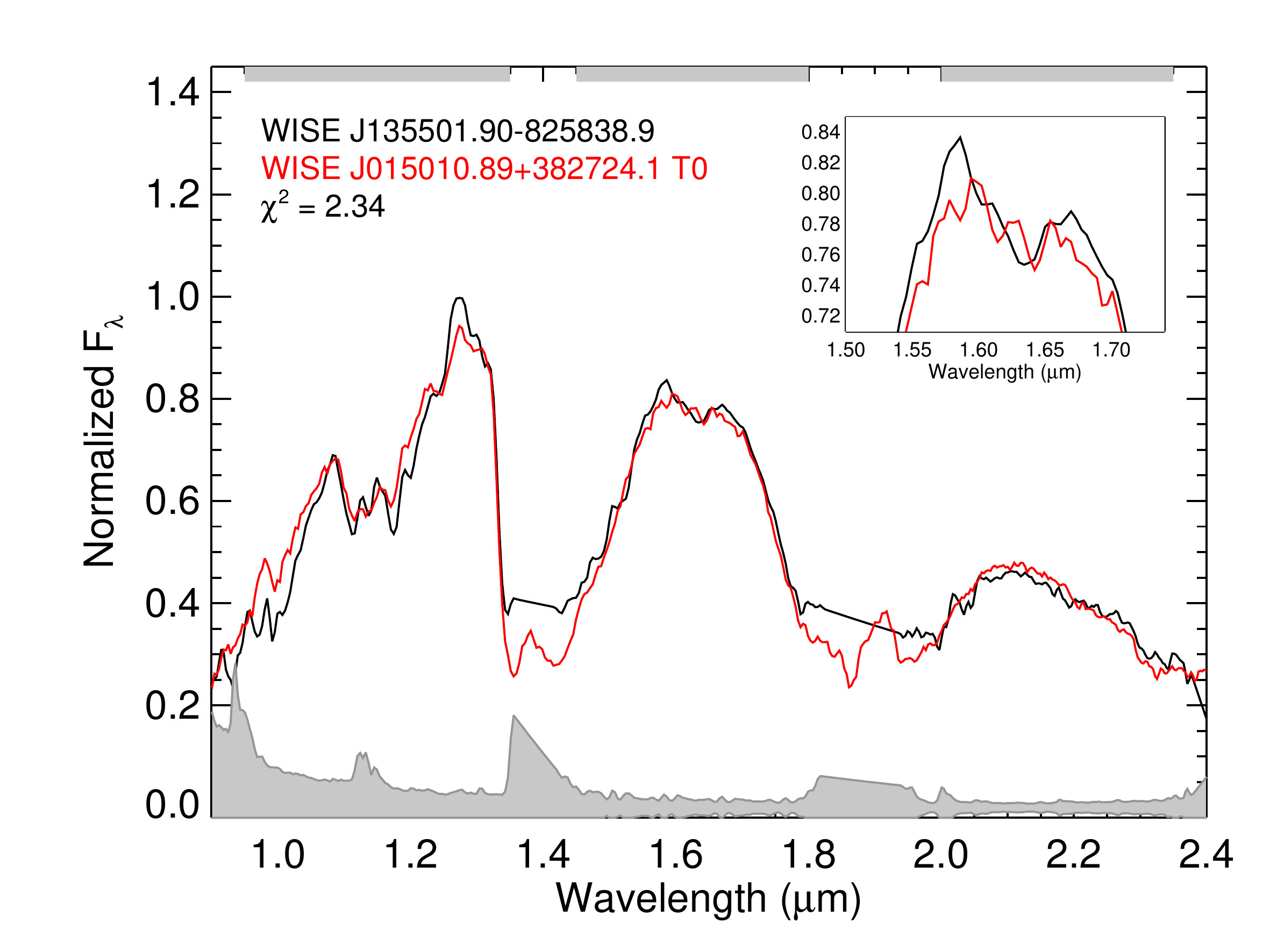}{0.5\textwidth}{(a) Best single fit for field L and T dwarf templates.\label{fig:singlefit}}
          \fig{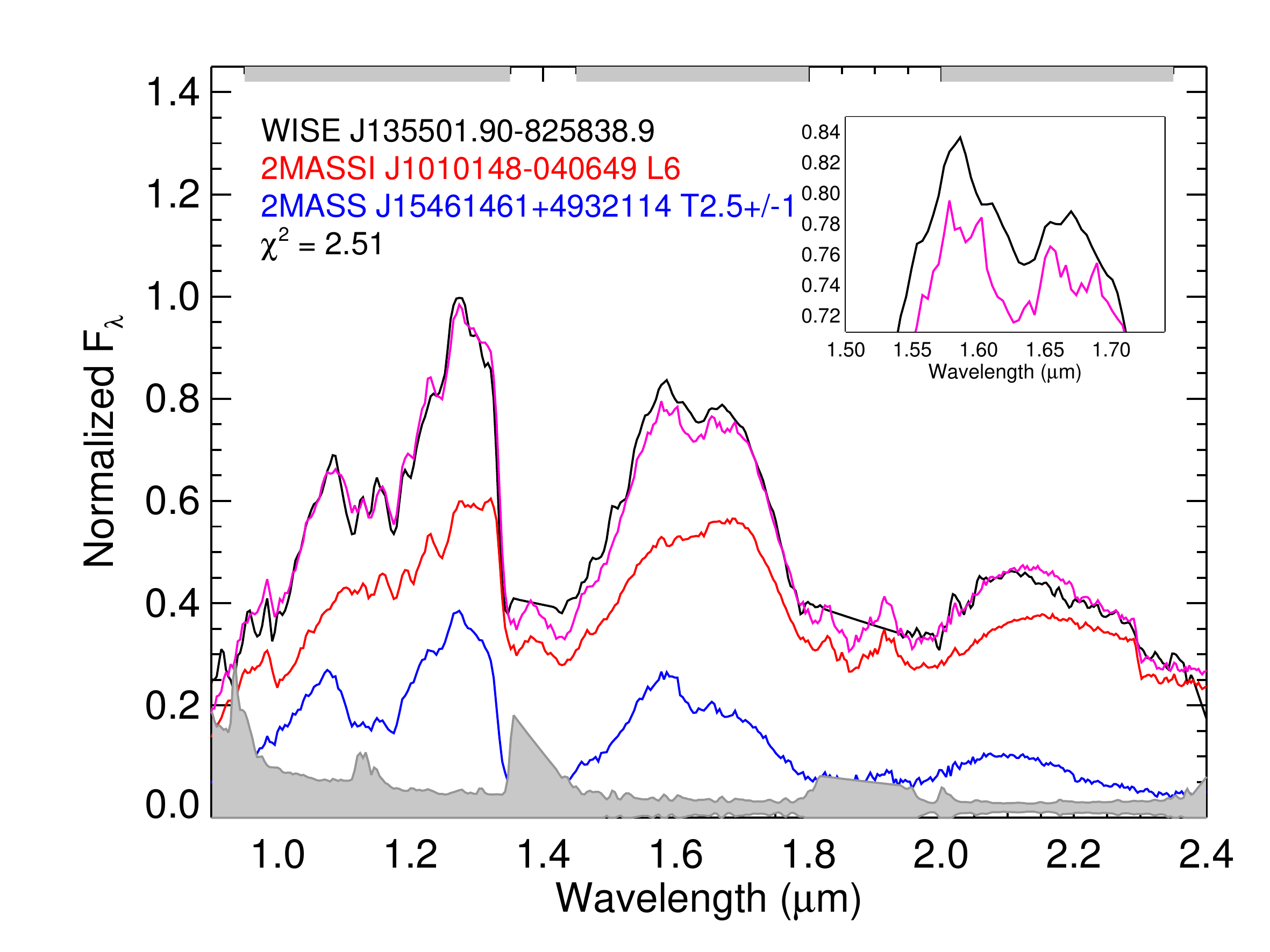}{0.5\textwidth}{(b) Best binary fit solution for field templates constrained to L0$-$L8 for primaries and T2$-$T9 for secondaries.\label{fig:T3binaryfit}}}
\gridline{\fig{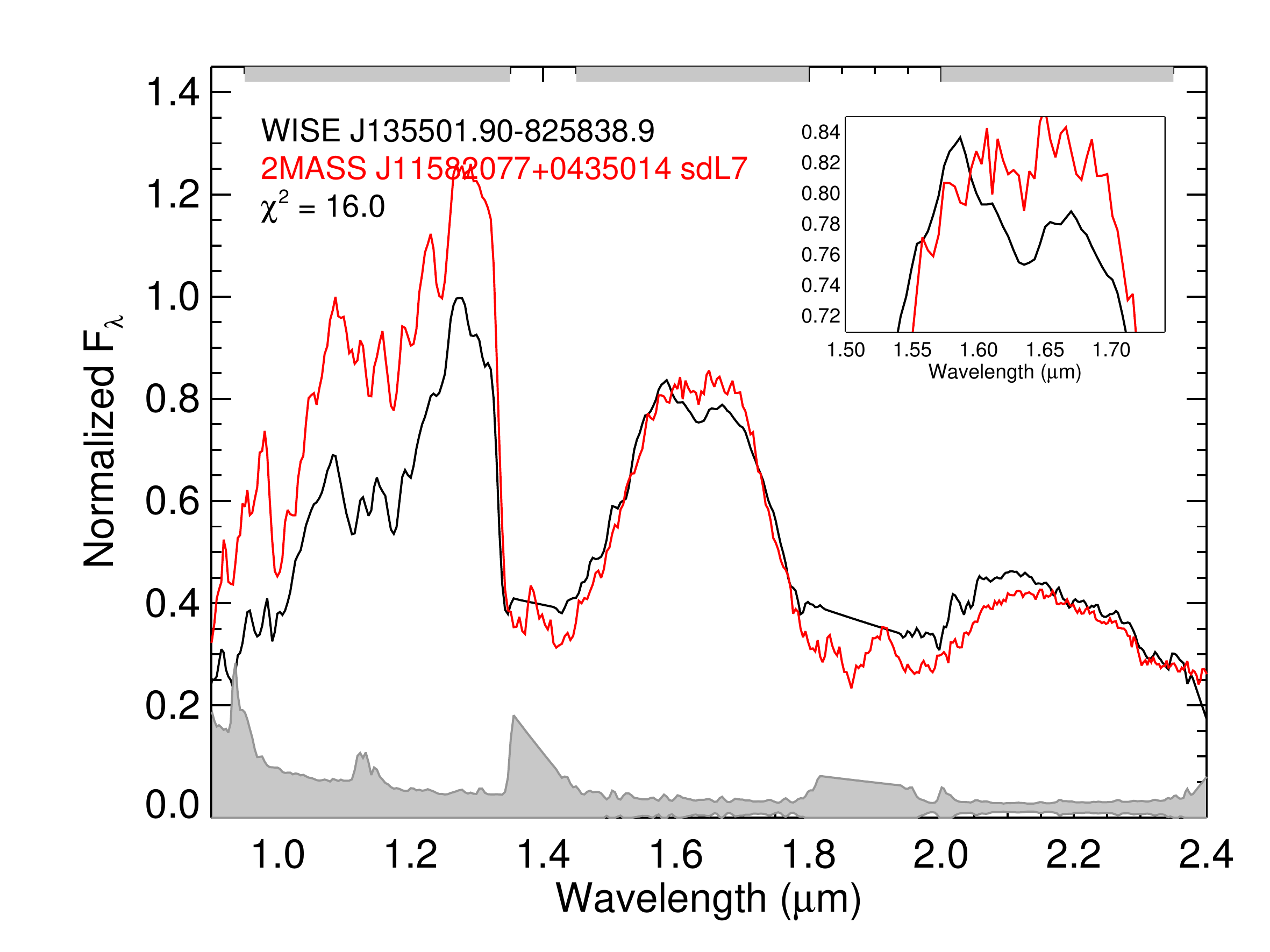}{0.5\textwidth}{(c) Best single fit using only L-type subdwarf templates.\label{fig:subdwarffit}}
	\fig{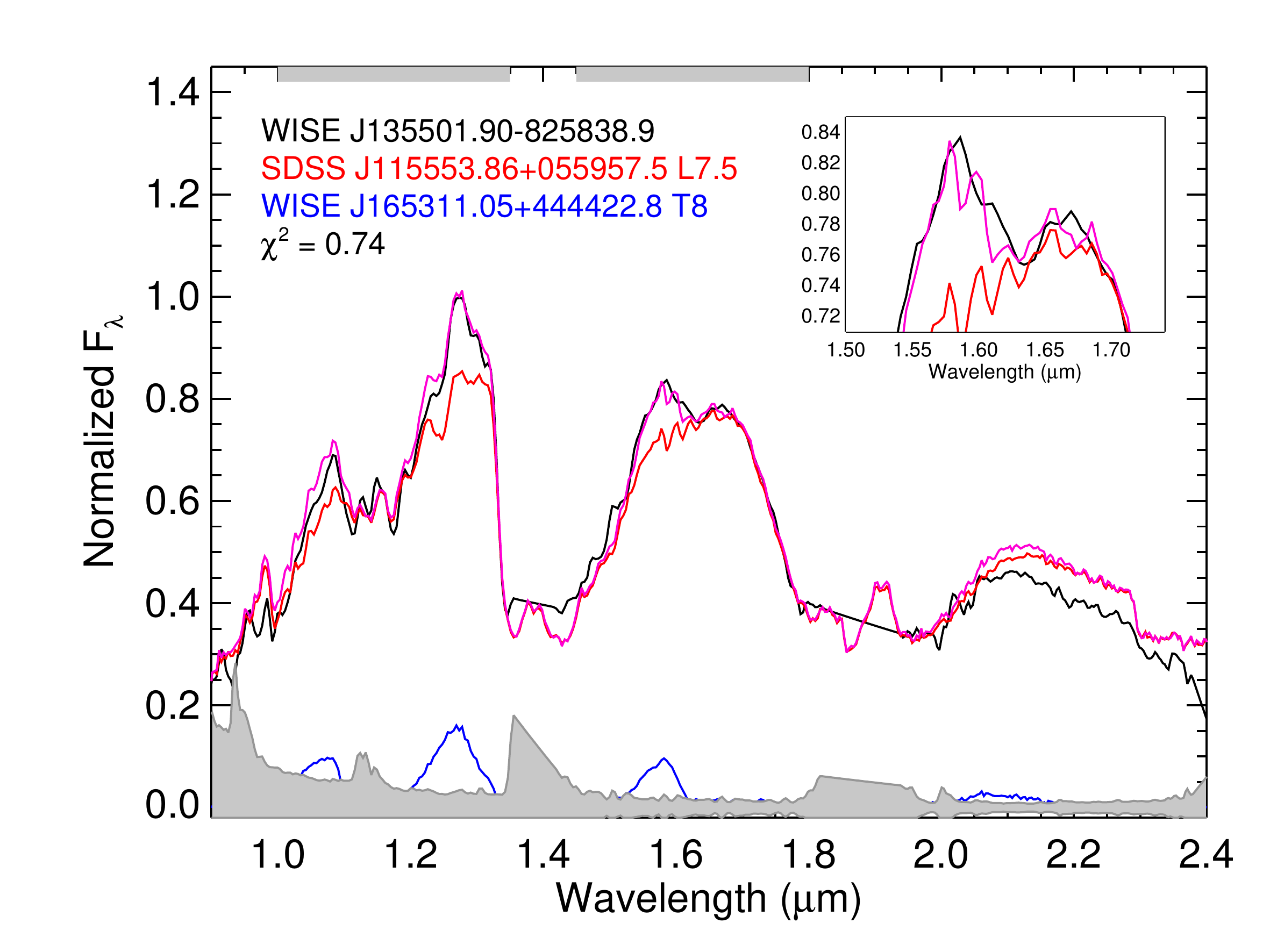}{0.5\textwidth}{(d) Best fit binary to the $J$ and $H$ bands only using field templates constrained to L0$-$L8 for the primary and T5$-$T9.5 templates for the secondary.\label{fig:JHT8binaryfit}}}
\caption{Best single and binary fits for WISE J1355$-$8258 from various template subsamples. The spectrum of the source (black) is normalized to the peak flux in the $1.0-1.3\,\mu$m range. Single fit templates are shown in red. Binary fits (magenta) are built by adding primary (red) and secondary (blue) templates scaled by their absolute magnitudes according to the $J$-band field relation of~\citet{2016ApJS..225...10F}. The methane dip feature at 1.63\,$\mu$m is seen in the inset. Wavelength regions used for the fitting are noted as bars shaded in grey at the top of the plots. Noise spectrum is shaded in grey.\label{fig:binfit}}
\end{figure*}

The reduced prism spectrum of WISE~J1355$-$8258 (Figure~\ref{fig:binfit}) displays strong H$_2$O absorption between $1.10-1.25\,\mu$m, $1.30-1.50\,\mu$m, and $1.80-2.00\,\mu$m, and visible methane absorption at $1.63\,\mu$m, both indicative of a T-dwarf. We determined formal near-infrared classifications of L9$\pm0.9$ using the indices of~\citet{2006ApJ...637.1067B} and T0$\pm$0.5 by direct comparison to the spectral standards of~\citet{2010ApJS..190..100K} using the SpeX Prism Library Analysis Toolkit (SPLAT;~\citealt{2014ASInC..11....7B}). These are later than both the~\citet{2016ApJ...817..112S} L7.1 photometric classification and the~\citet{2016ApJS..224...36K} sdL5$?$ spectral classification.  A comparison between WISE~J1355$-$8258 and a closely-matched T0 dwarf, WISE~J015010.89+382724.1~(\citealt{2011ApJS..197...19K}; Figure~\ref{fig:binfit}) reasonably fits the overall SED, but fails to match the absorption features in detail. Methane typically emerges at 2.2\,$\mu$m for L9 and at 1.6\,$\mu$m for T0~(\citealt{2002ApJ...564..466G,2006ApJ...639.1095B}), yet WISE~J1355$-$8258 shows the opposite trend: clear methane at 1.6\,$\mu$m but marginal at 2.2\,$\mu$m. The spectrum of WISE~J1355$-$8258 also has a slightly blue SED even after the photometric correction, which motivated the subdwarf designation of~\citet{2016ApJS..224...36K}. Both of these patterns have been previously observed in the blended-light spectra of L+T dwarf spectral binaries~(e.g.~\citealt{2004ApJ...604L..61C,2010ApJ...710.1142B}).

To assess the likelihood that WISE~J1355$-$8258 is an unresolved spectral binary, we followed the prescription of~\citet{2010ApJ...710.1142B}, comparing six sets of spectral indices to the loci of known L+T binaries. Index measurements of the blended-light spectrum of WISE~J1355$-$8258 satisfy four index-index selection criteria: H$_2$O$-J$ vs. H$_2$O$-K$, CH$_4-H$ vs. CH$_4-K$, CH$_4-H$ vs. $K/J$, and $H_2$O$-H$ vs. $H$-dip, thus qualifying it as a strong spectral binary candidate.


We compared the FIRE prism spectrum, sampled to a resolution of $\mathrm{R}\sim100$ to single and binary templates drawn from the SpeX Prism Library. We compared the spectrum to 1321 single templates in the SpeX Prism Library, excluding confirmed and candidate spectral binaries. We found a best match to the spectrum of the T0 WISE~J015010.89+382724.1~\citep{2011ApJS..197...19K} with a chi squared of $\chi^2 = 98.75$ or reduced chi squared $\chi^2_r=2.34$. This single template fails to reproduce the dip at $1.63\,\mu$m, or the slope of the peak at $1.27\,\mu$m, both typically encountered in spectral binaries with a T dwarf secondary.


We reexamined the hypothesis of~\citet{2016ApJS..224...36K}, that the peculiarities in the spectrum of WISE~J1355$-$8258 are due to subsolar metallicity. We compared the prism spectrum to eight L subdwarf spectra; Figure~\ref{fig:binfit} shows the best match as the sdL7 2MASS~J11582077+0435014~\citep{2009A&A...497..619Z}, with $\chi^2 = 675.2$ ($\chi^2_r=16.0$). This template is a poor match to the spectrum of WISE~J1355$-$8258, failing to reproduce the absorption feature at $1.63\,\mu$m and displaying a bluer SED. Despite our small comparison subsample, low metallicity is not a compelling explanation for this peculiar spectrum.

The spectrum of WISE~J1355$-$8258 was then compared to 81920 binary templates constructed from 640 L0$-$L8 primary and 128 T2$-$T9 secondary spectra, scaled by their absolute magnitudes in $J$, $H$ and $K_S$ bands according to~\citet{2016ApJS..225...10F}. We excluded templates with spectral types between L9-T1 to constrain the origin of the methane features to the T dwarf secondary, since T0 and T1 secondaries did not reproduce the $H$-band dip. The best match (Figure~\ref{fig:binfit}) is a combination of the L6 2MASSI~J1010148$-$040649~\citep{2003AJ....126.2421C} and the T2.5$\pm$1 2MASS J15461461+4932114~\citep{2008ApJ...676.1281M} yielding a $\chi^2 = 105.9$ ($\chi^2_R=2.51$). Despite the slightly higher $\chi^2$ than the best single fit, this binary template matches the spectral shape  of WISE~J1355$-$8258 in the $J$ band, and broadly reproduces the absorption features in the $H$ and $K_S$ bands, albeit with a slight shift in slope. 
After ranking and weighting all the binary fits by their $\chi^2$ we arrive at average component spectral types of L6.0$\pm$1.0 and T3.0$\pm$1.8.

This peculiar spectrum is unlike previous spectral binary candidates in that we cannot find a binary template that closely reproduces the absorption features and spectral slope simultaneously. This may be due to either residual spectral calibration issues (Section~\ref{sec:obsfire}) or surface gravity effects if the source is young (Section~\ref{sec:young}). In an effort to improve the fit of the peculiar absorption features of WISE~J1355$-$8258 regardless of its slope, we tried to fit binary templates to only the $J$ and $H$ bands of the spectrum. Most of the spectral binary indicators are in the $J$ and $H$ bands, and the $K_S$ band contributes less flux  than $J$ or $H$ to the overall spectrum. With this constraint, we obtain a significantly better fit. Shown in Figure~\ref{fig:binfit}, the best fit binary template is composed of the L7.5 SDSS~J115553.86+055957.5~\citep{2004AJ....127.3553K} and the T8 WISE~J165311.05+444422.8~\citep{2011ApJS..197...19K} with a $\chi^2 = 19.54$ ($\chi_R^2 = 0.74$). This binary template closely reproduces the peculiar features of WISE~J1355$-$8258 in the $J$ and $H$ bands, and the spectral shape of the $K_S$ band, except for the enhanced brightness. While the T8 secondary contributes only a small fraction of the total flux, it adds flux precisely in the spectral regions where the primary alone would miss the spectrum of WISE~J1355$-$8258. On average, the resulting component spectral types weighted by binary fit $\chi^2$ are L7.0$\pm$0.6 and T7.5$\pm$0.4.


In summary, we find that the spectrum of WISE J1355$-$8258 can be equivalently matched to both single and binary templates, although the latter more accurately reproduce detailed features associated with CH$_4$ absorption at 1.3\,$\mu$m and 1.6\,$\mu$m. We thus identify this source as a candidate substellar binary, and reject low metallicity as an explanation for its unusual near-infrared spectrum. Given that two binary combinations fit the spectrum, we consider the cases of L6+T3 and L7+T7.5 components separately.

\section{Kinematics and youth}\label{sec:young}

In the BASS-UC survey, potentially young sources are assigned a moving group membership probability that determines their priority for follow-up. The location of WISE~J1355$-$8258 in a $J$ versus $J-K$ color-magnitude diagram ($J = 16.14$, $J-K = 1.42\pm0.19$) is consistent with an L9 dwarf at the AB~Doradus kinematic distance. Additionally, given its red $W1-W2$ color ($W1-W2 = 0.57\pm0.04$), WISE~J1355$-$8258 became a high priority target for radial velocity (RV) measurements.

\subsection{Radial velocity measurements}
Two RV measurements were obtained from the FIRE echelle spectrum of \citet{2016ApJS..224...36K} on UT 2015 May 31 and the one we observed in UT 2017 June 11 (Section~\ref{sec:obsfire}). The RVs of both spectra were measured by comparing them with zero-velocity CIFIST 2011 BT-Settl models~(\citealt{2012EAS....57....3A,2015A&A...577A..42B}) using the IDL implementation of the Nelder-Mead downhill simplex method~\citep{doi:10.1093/comjnl/7.4.308} to minimize the $\chi^2$ residuals between the model and data. The free parameters were RV, width of the instrumental line spread function (assumed Gaussian), and a linear polynomial to account for slope systematics in the observed spectrum. The fitting procedure was applied in the $1.5100-1.5535\,\mu$m region of the $H$-band, where the signal-to-noise ratio and absorption feature density are high in L dwarfs. The wavelength range was divided in fifteen evenly-distributed 0.02\,$\mu$m-wide regions to perform model fits. The two spectra yielded respective RVs of  $20 \pm 5$\,\kms\ and $26 \pm 6$\,\kms, each including a systematic uncertainty of $\pm 3$\,\kms\ \citep{2017ApJS..228...18G}. The center of mass of the system shows a stable RV over measurements separated by a 2-year period, supporting its AB Doradus membership. However, more epochs of more precise RV are needed to confirm or reject the AB Doradus membership, and to determine whether the binary motion is convolved with the young moving group proper motion. We adopt their inverse variance-weighted average of $22 \pm 5$\,\kms as the systemic RV.

Combining sky position, proper motion from 2MASS to AllWISE~\citep{2016ApJ...817..112S}, and RV, BANYAN~II gives a Bayesian probability of $95.6\%$ for AB Doradus membership (without assuming a young age), with a $4.0\%$ field contamination probability. BANYAN~II yields a kinematic distance of $17 \pm 2$\,pc. Based on the proper motion, RV and the kinematic distance estimate for the young L7+T7.5 case, the $UVW$ velocities for this object are: $U = -7 \pm 4$\,\kms, $V = -27 \pm 4$\,\kms, and $W = -13 \pm 2$\,\kms, placing it at $\sim$\,0.8\,\kms and 21\,pc from the kinematic and spatial centers of the BANYAN~II AB~Doradus models. Conversely, if the object is a field binary, using the spectrophotometric distances, we arrive at $U = -25\pm$9\,\kms, $V = -38 \pm 8$\,\kms, and $W = -19 \pm 7$\,\kms for the field L6+T3 case and $U = -18\pm$4\,\kms, $V = -34 \pm 4$\,\kms, and $W = -17 \pm 3$\,\kms for the field L7+T7.5 case.

\begin{deluxetable*}{llcccc}
\tabletypesize{\footnotesize}
\tablecolumns{6}
\tablewidth{0pt}
\tablecaption{Equivalent widths of K I doublets for WISE 1355$-$8258 compared to field and young single and synthetic binary templates. \label{tab:KI}}
\tablehead{
\nocolhead{} &
\colhead{Source} & 
\colhead{1.169$\mu$m} & 
\colhead{1.177$\mu$m} & 
\colhead{1.243$\mu$m} & 
\colhead{1.254$\mu$m}}
\startdata
\hline
Combined epochs & WISE J135501.90$-$825838.9 & 7.1$\pm$0.6  & 11.1$\pm$0.6  & 3.9$\pm$0.3 & 6.0$\pm$0.3 \\
 \hline
 \hline
\sidehead{\emph{Case 1: Field L6+T3}}
Field L6 & 2MASS J08503593+1057156 & 6.4$\pm$1.5 & 9.1$\pm$2.0 & 4.6$\pm$0.9 & 5.8$\pm$1.2 \\ 
Field T3.5 & SDSSp J175032.96+175903.9 & 9.8$\pm$1.8 & 13.7$\pm$2.2 & 5.6$\pm$0.9 & 9.8$\pm$1.1   \\
Field L6+T3.5 & Synthetic Binary &  7.8$\pm$1.5 & 10.9$\pm$2.0 & 5.1$\pm$0.9 & 7.6$\pm$1.1\tablenotemark{*}  \\
\hline
\hline
\sidehead{\emph{Case 2: Field L7+T7.5}}
Field L7 & 2MASS J17281150+3948593 & 7.5$\pm$1.9 & 10.1$\pm$2.1 & 5.9$\pm$1.3 & 7.1$\pm$1.3   \\
Field T7 & 2MASS J07271824+1710012 & 8.9$\pm$2.7 & 11.1$\pm$2.7 & 3.2$\pm$0.6 & 6.5$\pm$0.8   \\
 Field L7+T7 & Synthetic Binary  & 7.8$\pm$1.8  & 10.2$\pm$2.0 & 4.8$\pm$1.0 & 6.5$\pm$1.1   \\
\hline
\hline
\sidehead{\emph{Case 3: Young L7+T7.5}}
Young L6 & 2MASS J22443167+2043433 & 3.5$\pm$1.1 & 5.1$\pm$1.5 & 1.9$\pm$0.7 & 2.1$\pm$0.8  \\
Field T7 & 2MASS J07271824+1710012 & 8.9$\pm$2.7 & 11.1$\pm$2.7 & 3.2$\pm$0.6 & 6.5$\pm$0.8   \\
 Young L6 + Field T7 & Synthetic Binary & 3.8$\pm$1.1\tablenotemark{*}  & 5.4$\pm$1.6\tablenotemark{*}  & 2.1$\pm$0.6\tablenotemark{*}  & 3.0$\pm$0.7\tablenotemark{*} \\ 
 \hline
 \hline
\vspace{-0.6cm}
\enddata
\tablenotetext{*}{Equivalent widths that are statistically different from those of WISE~J1355$-$8258.}
\tablecomments{Equivalent width averages are weighted by their inverse variances. In cases 2 and 3, a field T6.5 template was used to calculate K~I equivalent widths due to the lack of available T7.5 high resolution spectral templates. In case 3, a young L6 template and AB~Doradus candidate~\citep{2016ApJS..225...10F} was used instead of an L7 template for the same reason. NIRSPEC spectra for all the sources were obtained from~\citet{2003ApJ...596..561M}. 2MASS J0850+1057 and 2MASS J1728+3948 are both binary systems with near equal spectral types~\citep{2011AJ....141...70B}. A potential case 4 for a young L6+T3 binary is disproved in Section~\ref{sec:absmag}.}
\end{deluxetable*}%

 \subsection{Gravity-sensitive indicators}
The temperature and gravity-sensitive K~I doublets at $1.17\,\mu$m and $1.25\,\mu$m can provide an independent indication of youth. Lower surface gravity brown dwarfs exhibit narrower alkali lines and weaker equivalent widths compared to field-age objects due to their lower atmospheric pressure\footnote{Young brown dwarfs (ages $\lesssim$ 200\,Myr;~\citealt{2001RvMP...73..719B}) have lower surface gravity than equivalent-temperature field dwarfs due to their inflated radii and lower masses at equivalent luminosities.}~\citep{2004ApJ...600.1020M,2017ApJ...838...73M}. 

Table~\ref{tab:KI} compares K~I equivalent widths for WISE~J1355$-$8258 FIRE echelle data and similarly-classified single sources and combined-light synthetic binaries. The primaries of these binaries are selected to be representative field age sources (2MASS J0850+1057 and 2MASS J1728+3948) and AB Doradus members (2MASS~J2244+2043;~\citealt{2016ApJS..225...10F}). The T dwarf secondaries are all field sources. The individual measurements were performed on a combined spectrum of the~\citet{2016ApJS..224...36K} and our FIRE echelle spectra by an inverse variance-weighted average. The K~I equivalent widths of WISE~J1355$-$8258 and those from both field cases are statistically consistent.  Finally, the equivalent widths of WISE~J1355$-$8258 are all at least $1\sigma$ higher than those for the case of the synthetic binary composed of a young L7 and field T7.5 objects. 

These comparisons suggest that this peculiar spectrum has gravity indicators more consistent with a field binary. However, we note that in all three cases, the potassium equivalent widths of the synthetic binaries are stronger than those of their respective primary components, yet the effect is more significant in the field cases. We note that since most of the templates in the SpeX Prism Library have field ages, the relative magnitudes presented for both combinations of spectral types are based on field templates of L7 and T7.5.  The relative magnitudes affect the relative scaling of the synthetic binaries. 

\begin{deluxetable*}{lcccccccc}
\tablecolumns{9}
\tabletypesize{\footnotesize}
\tablewidth{0pt}
 \tablecaption{Properties of the WISE J1355$-$8258 system.\label{tab:binprops}} 
 \tablehead{
 & & \multicolumn{2}{c}{Case 1}
 & \multicolumn{2}{c}{Case 2}
 & \multicolumn{2}{c}{Case 3}\\
 \cline{3-8}
 \colhead{Property} &
 \colhead{System} &
 \colhead{A} & 
 \colhead{B} & 
 \colhead{A} & 
 \colhead{B} & 
 \colhead{A} & 
 \colhead{B} & 
 \colhead{Reference}}
\startdata
NIR Spectral Type & L9pec & L6.0$\pm$1.0 & T3.0$\pm$1.8 & L7.0$\pm$0.6 & T7.5$\pm$0.4 & L7.0$\pm$0.6 & T7.5$\pm$0.4 &  1\\
Assumed Age (Gyr) & \nodata & \multicolumn{2}{c}{$2-5$} & \multicolumn{2}{c}{$2-5$} & \multicolumn{2}{c}{$0.13-0.2$} & 1\\
\hline
\sidehead{\emph{Magnitudes}}
$\Delta$ 2MASS $J$\tablenotemark{$\star$}	& \nodata & \multicolumn{2}{c}{1.0$\pm$0.8} & \multicolumn{2}{c}{2.4$\pm$0.4} & \multicolumn{2}{c}{2.4$\pm$0.4} & 1\\ 
$\Delta$ 2MASS $H$\tablenotemark{$\star$}  & \nodata & \multicolumn{2}{c}{1.4$\pm$0.9} & \multicolumn{2}{c}{3.5$\pm$0.3} & \multicolumn{2}{c}{3.5$\pm$0.3} & 1\\
$\Delta$ 2MASS $K_s$\tablenotemark{$\star$} & \nodata & \multicolumn{2}{c}{1.9$\pm$1.0} & \multicolumn{2}{c}{4.1$\pm$0.4} & \multicolumn{2}{c}{4.1$\pm$0.4}  & 1\\
2MASS $J$ & 16.14$\pm$0.13 & 16.5$\pm$0.3 & 17.5$\pm$0.6 & \nodata & \nodata & 16.3$\pm$0.1 & 18.6$\pm$0.4 & 1,2\\
2MASS $H$ & 15.31$\pm$0.13 & 15.6$\pm$0.3 & 17.0$\pm$0.8 & \nodata & \nodata & 15.4$\pm$0.1 & 18.9$\pm$0.3 & 1,2\\ 
2MASS $K_S$ & 14.72$\pm$0.14 & 14.9$\pm$0.2 & 16.8$\pm$0.8 & \nodata & \nodata & 14.7$\pm$0.1 & 18.9$\pm$0.4 & 2\\ 
FourStar $J$ & 16.47$\pm$0.05 & \nodata & \nodata & \nodata & \nodata & \nodata & \nodata &  1\\
FourStar $H$ & 15.45$\pm$0.04 & \nodata & \nodata & \nodata & \nodata & \nodata & \nodata &  1\\
FourStar $K_S$ & 15.01$\pm$0.05 & \nodata & \nodata & \nodata & \nodata & \nodata & \nodata &  1\\
WISE $W1$ & 14.12$\pm$0.03 & \nodata & \nodata & \nodata & \nodata & \nodata & \nodata & 2\\ 
WISE $W2$ & 13.55$\pm$0.03 & \nodata & \nodata & \nodata  & \nodata & \nodata & \nodata &  2\\ 
WISE $W3$ & 12.5$\pm$0.3 & \nodata & \nodata & \nodata  & \nodata & \nodata & \nodata & 2\\ 
WISE $W4$ & $\leq$ 9.7 & & \nodata & \nodata \nodata  & \nodata & \nodata & \nodata & 2\\
M$_J$\tablenotemark{$*$} & 15.0$\pm$0.4, 15.1$\pm$0.4 & 14.2$\pm$0.4 & 14.8$\pm$0.4 & \nodata & \nodata & 15.1$\pm$0.4 & 17.5$\pm$0.6 & 1\\
M$_H$\tablenotemark{$*$} & 14.1$\pm$0.4, 13.8$\pm$0.4 & 13.0$\pm$0.4 & 14.1$\pm$0.4 & \nodata & \nodata & 14.2$\pm$0.4 & 17.7$\pm$0.6 & 1\\
M$_{K_S}$\tablenotemark{$*$} & 13.5$\pm$0.4, 13.2$\pm$0.4 & 12.3$\pm$0.4 & 14.0$\pm$0.4 & \nodata & \nodata & 13.6$\pm$0.4 & 17.7$\pm$0.7 & 1\\
\hline
\sidehead{\emph{Kinematics}}
RV (\kms) & 22$\pm$5 & \nodata & \nodata & \nodata & \nodata & \nodata & \nodata & 1\\ 
$\mu_\alpha\cos\delta$ (mas\,yr$^{-1}$) & -241$\pm$8 & \nodata & \nodata & \nodata & \nodata & \nodata & \nodata & 2\\
$\mu_\delta$ (mas\,yr$^{-1}$) & -142$\pm$14 & \nodata & \nodata & \nodata & \nodata & \nodata & \nodata & 2\\
$d\dagger$ (pc) & \nodata & 33$\pm$9 & 33$\pm$19 & 27$\pm$3 & 27$\pm$4 & \multicolumn{2}{c}{17$\pm$2} & 1\\
$U$\tablenotemark{$\ddagger$} (\kms) & \nodata & \multicolumn{2}{c}{-25$\pm$9} & \multicolumn{2}{c}{-18$\pm$4} &\multicolumn{2}{c}{-7$\pm$4} & 1\\
$V$\tablenotemark{$\ddagger$} (\kms) & \nodata & \multicolumn{2}{c}{-38$\pm$8} & \multicolumn{2}{c}{-34$\pm$4} & \multicolumn{2}{c}{-27$\pm$4} & 1\\
$W$\tablenotemark{$\ddagger$} (\kms) & \nodata & \multicolumn{2}{c}{-19$\pm$7} & \multicolumn{2}{c}{-17$\pm$3} & \multicolumn{2}{c}{-13$\pm$2} & 1\\
\hline
\sidehead{\emph{Masses}}
Mass (\mjup) & \nodata & $72^{+4}_{-5}$ & $61^{+6}_{-8}$ & $70^{+2}_{-4}$ & $42^{+5}_{-6}$ & 11$\pm$1 & 9$\pm$1 & 1\\ 
Mass ratio & \nodata & \multicolumn{2}{c}{0.84$\pm$0.06} & \multicolumn{2}{c}{0.60$\pm$0.08} & \multicolumn{2}{c}{0.82$\pm$0.02} & 1\\ 
\enddata
\tablerefs{(1) This paper; (2)~\citealt{2016ApJ...817..112S}; (3)~\citealt{2015MNRAS.454..593B}.}
\tablenotetext{\star}{The magnitude differences presented here come from the spectral binary fitting. Since the templates used in the fitting are primarily of field age, we are limited to use the same magnitude differences for the young case.}
\tablenotetext{$*$}{Combined-light absolute magnitude for young and field assumptions, respectively.}
\tablenotetext{\dagger}{Spectrophotometric distances for Case 1 and Case 2; kinematic distance for Case 3.}
\tablenotetext{\ddagger}{$UVW$ velocities based on spectrophotometric distance for Cases 1 and 2; kinematic distance for Case 3.}
\end{deluxetable*}

\section{Analysis}\label{sec:analysis}

\subsection{Spectrophotometric distances}

The binary fitting procedure yields the relative magnitudes between primary and secondary in 2MASS $J$, $H$ and $K_S$ filters, which along with the combined magnitude, can be used to find individual component magnitudes. In order to apply them to our FourStar $J$, $H$ and $K_S$-band measurements, we investigated the difference in the two photometric filter systems from synthetic magnitude measurements on the SpeX spectra of the best-fitting binary template (Figure~\ref{fig:binfit}). The filter profiles\footnote{Available at the SVO filter profile service \url{http://svo2.cab.inta-csic.es/svo/theory/fps/}.} of each band in the 2MASS and FourStar systems were convolved on the spectra of the observed brown dwarf template and Vega to obtain the synthetic magnitudes. We found that additive offsets of only 0.03, $-0.01$ and $-0.04$\,mag are needed to transform the relative 2MASS magnitudes to the FourStar system. Since these offsets were smaller than the uncertainties on the relative magnitudes, we used the 2MASS magnitude differences directly on the FourStar photometry to determine individual apparent magnitudes.

The spectrophotometric distances of the WISE~J1355$-$8258 system were estimated by calculating a distance modulus per filter with the apparent FourStar $JHK_S$ magnitudes and the corresponding 2MASS absolute magnitudes from the field relations of~\citet{2016ApJS..225...10F}. Distances were obtained for each component and filter, and combined per component with an average weighted by the inverse variance. The combined uncertainty was set to be equal to the smallest uncertainty in the individual filter estimates due to the correlation between apparent magnitude measurements. The resulting distances for the components of the field L6+T3 case are 33$\pm$9\,pc and 33$\pm$19\,pc for primary and secondary, respectively (See Table~\ref{tab:binprops}). For the field L7+T7.5 case, the component spectrophotometric distances are 27$\pm$3\,pc and 27$\pm$4\,pc. 

\subsection{Absolute Magnitude Diagrams}\label{sec:absmag}

\begin{figure*}
	\centering
	\subfigure[2MASS $J$-band]{\includegraphics[width=0.5\textwidth]{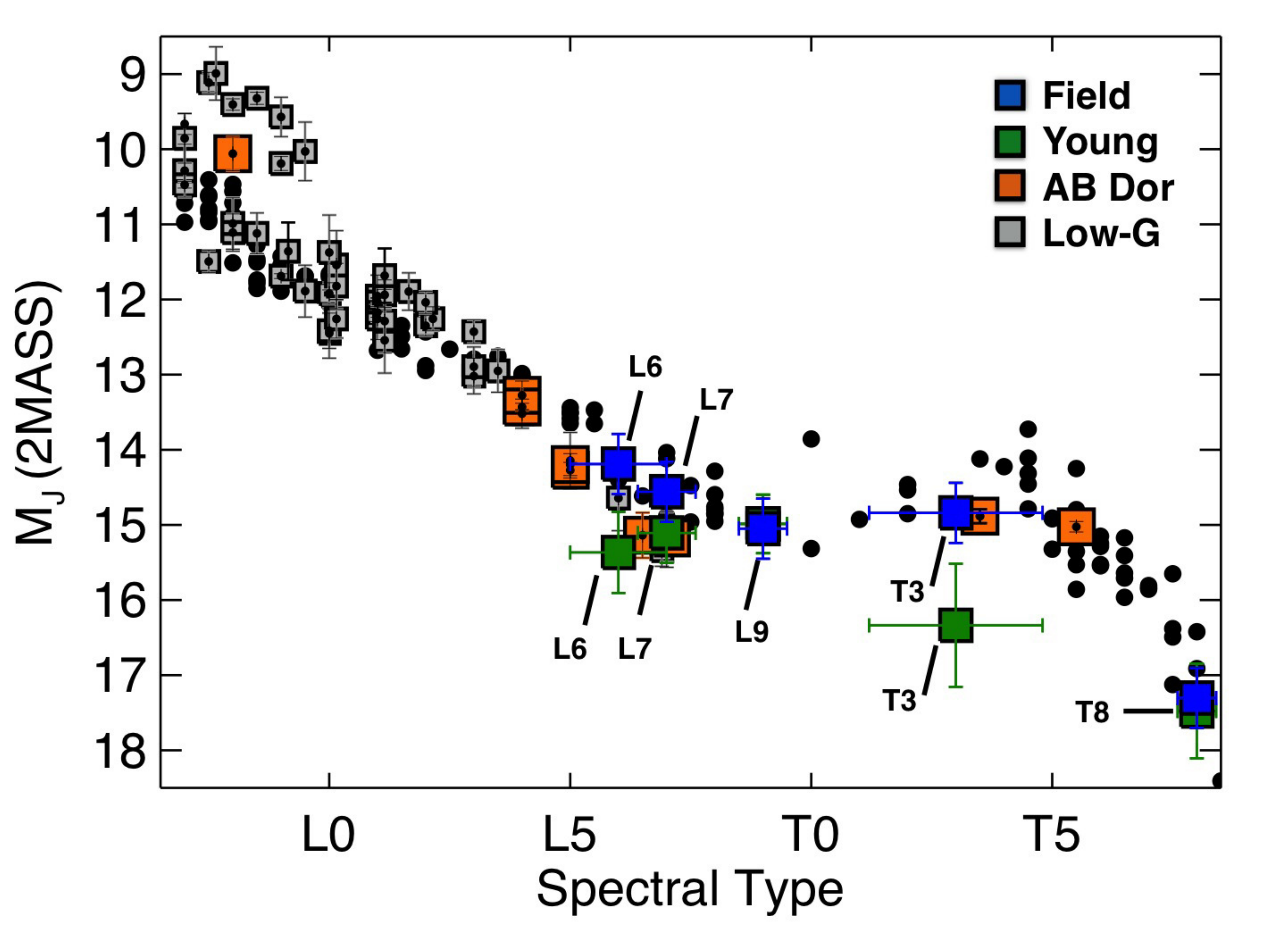}\label{fig:absmagJ}}\\
	\subfigure[2MASS $H$-band]{\includegraphics[width=0.5\textwidth]{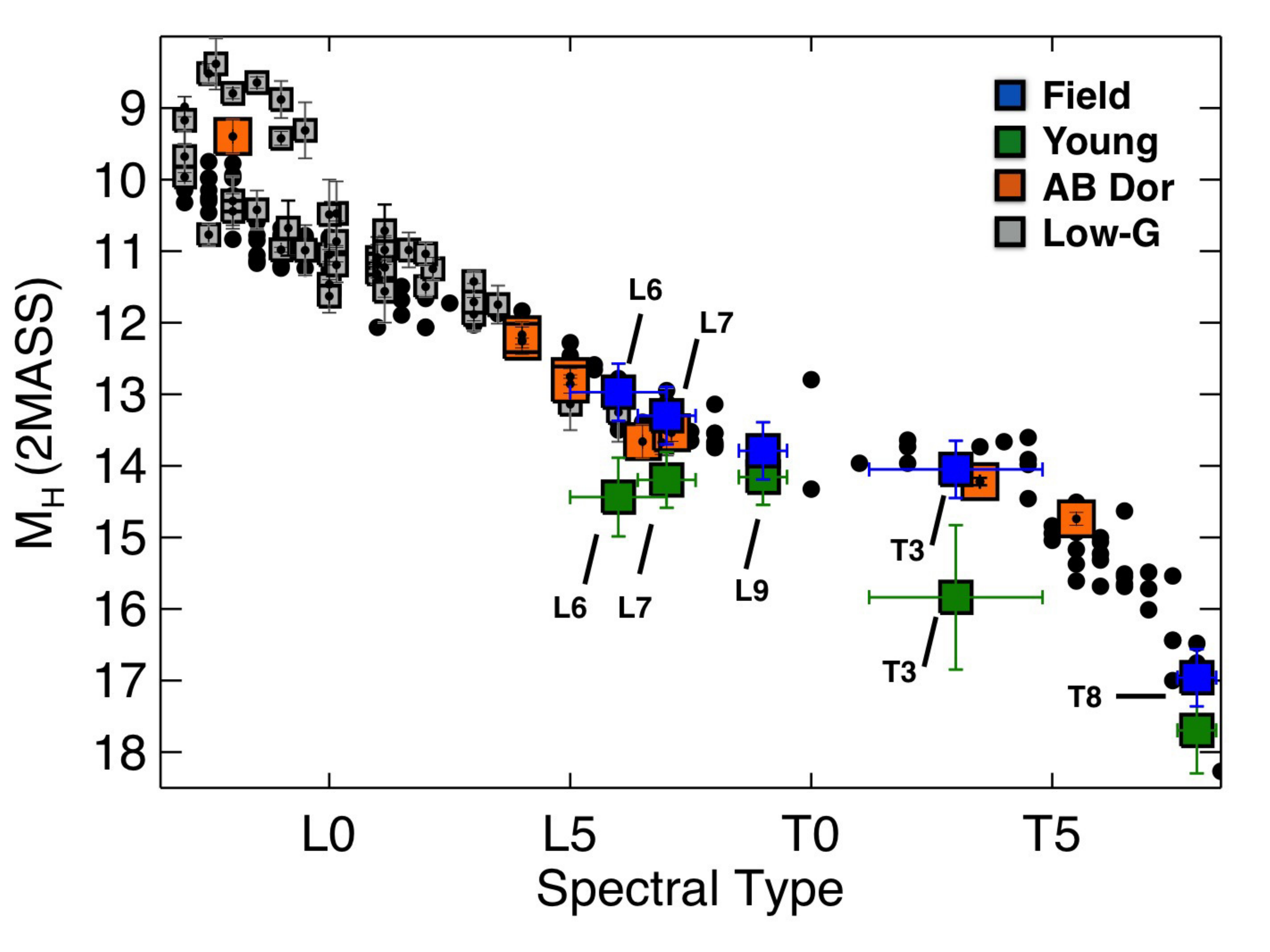}\label{fig:absmagH}}\\
	\subfigure[2MASS $K_S$-band]{\includegraphics[width=0.5\textwidth]{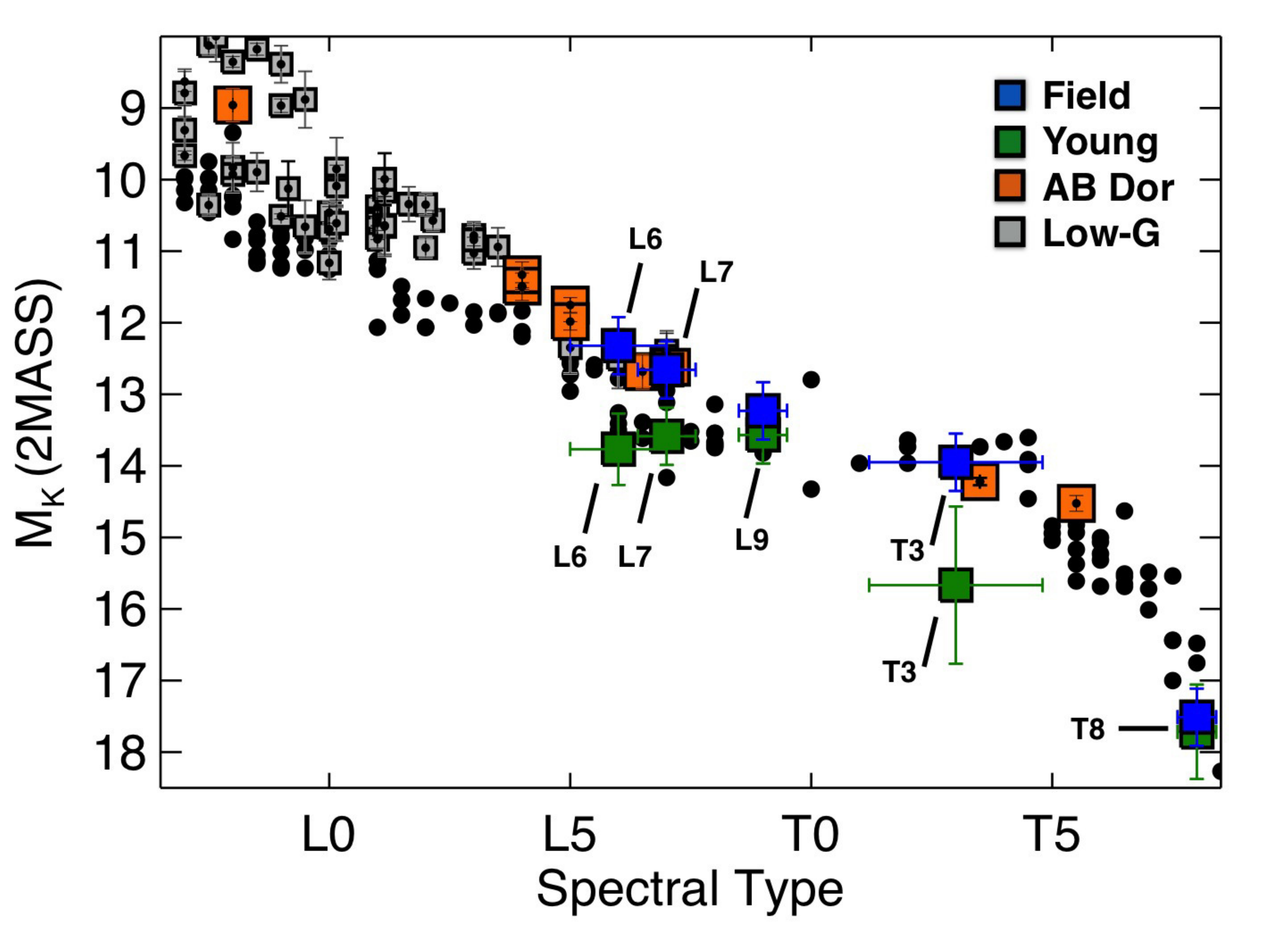}\label{fig:absmagK}}\\
\caption{Absolute magnitude diagrams showing the combined spectral type and individual binary components assuming the kinematic distance if young, and their absolute magnitude estimated from the field~\citet{2016ApJS..225...10F} relation. Boxes in green reflect the young assumption, and boxes in blue show the field assumption. Data points for field objects are shown as black dots~\citep{2016ApJS..225...10F}, low gravity objects are enclosed in grey boxes, and AB Doradus members are enclosed in orange boxes. Both components in the young L6+T3 case are underluminous compared to both field and young sequences, therefore, they are dropped from the analysis. \label{fig:absmag}}
\end{figure*}

We placed WISE~J1355$-$8258 on spectral type versus absolute magnitude diagrams as a combined-light object, and separated into both cases of spectral type components  (Figure~\ref{fig:absmag}). Additionally, since there is a slight possibility of AB Doradus membership, we considered both young and field ages. Absolute magnitudes were calculated assuming the kinematic distance for a young age case, and alternatively, using the field spectral type to absolute magnitude relations of~\citet{2016ApJS..225...10F} for the field case. This breakdown provided four cases: field L6+T3, young L6+T3, field L7+T7.5 and young L7+T7.5.

The young L6+T3 case produced absolute magnitudes at the kinematic distance over 1 magnitude underluminous for each spectral type compared to both field and young sequences. The unphysical placement of these components on the diagram disproves this case as a possible solution.  A more intriguing possibility is that WISE~J1355$-$8258 is a young spectral binary system of L7 and T7.5 components.  Using the kinematic distance, both components of this binary system are well placed on the $J$-band absolute magnitude diagram (albeit slightly underluminous on the $H$-band), providing independent support for the youth hypothesis. Both field cases are shown in Figure~\ref{fig:absmag} for completeness, fixed to the sequence by the absolute magnitude relation. A parallax measurement will be crucial to determine the identity of this source.

\subsection{Component Masses}
The component masses of WISE~J1355$-$8258 were estimated using the 2MASS isochrones of BT-Settl models \citep{2003A&A...402..701B, 2012RSPTA.370.2765A}, as described in~\citet{2015ApJS..219...33G} with Monte Carlo methods. Ages were drawn from a log-uniform distribution, while absolute magnitudes per filter were drawn from a normal distribution. Constrained by the relative magnitudes between components, the magnitudes were placed on isochrones to determine probability density functions of mass per filter per component. The mass estimates per filter were combined into one estimate per component in a similar manner to the weighted average of spectrophotometric distances. This process was followed for both young ($130-200$\,Myr if AB~Doradus member) and field ($2-5$\,Gyr) age assumptions, and corresponding absolute magnitudes based on the kinematic ($d_{kin}=17 \pm 2$\,pc) and weighted averages of the spectrophotometric ($d_{phot} = 33 \pm 8$\,pc and $d_{phot} = 27 \pm 3$\,pc for L6+T3 and L7+T7.5 cases) distances respectively.  The resulting values are reported in Table~\ref{tab:binprops}. The error bars only include uncertainty on the age, distance and relative magnitudes of WISE~J1355$-$8258, and do not include model systematics, which are not well constrained especially at young ages. The resulting masses for the L6+T3 field case are $72^{+4}_{-5}$\,\mjup~and $61^{+6}_{-8}$\,\mjup, and for the L7+T7.5 field case are $70^{+2}_{-4}$\,\mjup~and $42^{+5}_{-6}$\,\mjup. Interestingly, if the system is a young binary, its component masses would be 11$\pm$1\,\mjup~and 9$\pm$1\,\mjup, with both masses below the Deuterium burning limit. These masses lead to a mass ratio of $q\sim0.84\pm0.06$ for the field L6+T3 case, a fairly low $q \sim0.60\pm0.08$ for the field L7+T7.5 and $q \sim0.82\pm0.02$ for the young L7+T7.5 case. The field masses and subsequent mass ratios are based on a field age assumption of $2-5\,$Gyr, thus older ages would lower the masses and mass ratios.

\section{Discussion and Summary}\label{sec:discussion}

WISE~J1355$-$8258 is a source with a peculiar near-infrared, low-resolution spectrum that shows absorption features commonly found in blended-light spectral binaries. Through binary template fitting, we determined two spectral type compositions able to reproduce the peculiarities of this spectrum: L6+T3 and L7+T7.5. Independent measurements of the kinematics of this system used in BANYAN II yield a $95.6\%$ probability of membership in the AB~Doradus young moving group. Equivalent widths from K~I absorption lines acting as gravity-sensitive indicators in the spectrum of WISE~J1355$-$8258 most closely match those of a synthetic binary with field L6+T3 components. 

While the majority of templates used in the binary fitting have field ages, the high AB~Doradus membership probability prompted us to consider four cases: field L6+T3, young L6+T3, field L7+T7.5 and young L7+T7.5. However, the young L6+T3 case was dropped because its absolute magnitudes at the predicted kinematic distance yielded underluminous components. Therefore, if this system is young, the only viable possibility for component spectral types would be L7+T7.5. If this system has field age, both spectral combinations are possible explanations for the spectral peculiarities.

An important obstacle in trying to characterize this source, other than the current lack of a parallax measurement, is the sparse template sample of young brown dwarf spectra. While the SpeX Prism Library contains over 100 examples, there are not enough templates per spectral subtype nor young spectral type to absolute magnitude relations reliably extending into the T dwarf regime to reproduce the peculiarities introduced by low surface gravity and binarity simultaneously. 

If young, WISE~J1355$-$8258 would appear to be a binary composed of two planetary-mass objects, both below the Deuterium-burning minimum mass. The only other system somewhat comparable to WISE~J1355$-$8258 in terms of masses is the L7+L7 2MASS~J11193254$-$1137466AB~\citep{2016ApJ...821L..15K}, where both components have $3.7^{+1.2}_{-0.9}$\,\mjup~\citep{2017ApJ...843L...4B}~if they are members of TW~Hya~(10$\pm$3\,Myr;~\citealt{2015MNRAS.454..593B}). However, WISE~J1355$-$8258 would be the first instance of an L+T young spectral binary system. If confirmed, this will be one of a few age-calibrated, low-mass binary brown dwarfs with potentially measurable dynamical masses.  Otherwise, if this is a field spectral binary, then it is also an interloper to AB~Doradus, thus serving as a warning for future moving group kinematic studies that include binary systems. In either case, this is an L+T binary system in the Solar neighborhood with a peculiar spectrum, difficult to generate by current state-of-the-art formation simulations~(e.g.~\citealt{2012MNRAS.419.3115B}).

Regardless of whether WISE~J1355$-$8258 is a young binary, our analysis demonstrates the spectral binary method as a new technique for the discovery of planetary-mass companions to low-mass objects. However, a larger sample of young templates per spectral type is needed to accurately reproduce the peculiar absorption features caused by binarity along with the redder slopes of young objects. Searches for young T-dwarfs have been carried out with limited success in the past using methane imaging~(e.g.~\citealt{2009A&A...508..823B,2010ApJ...719L..90H,2013MNRAS.430.1208P}) and proper motions aligned with cluster membership~(e.g.~\citealt{2007MNRAS.378.1131C, 2007MNRAS.378L..24B}) which have returned a few candidates~(e.g.~\citealt{2007MNRAS.378L..24B,2016A&A...586A.157P}). However, there are only three verified young T dwarfs with fully measured kinematics: SDSS~J111010.01+011613.1~(T5.5pec;~\citealt{2015ApJ...808L..20G}), GU~Psc b~(T3.5$\pm$1;~\citealt{2014ApJ...787....5N}) and 51~Eri~b~(T6.5$\pm$1.5;~\citealt{2015Sci...350...64M,2017AJ....154...10R}). 

In summary, WISE~J1355$-$8258 is a strong spectral binary candidate. The only evidence we find for youth is its set of kinematic measurements that yield the possible AB~Doradus membership. Resolved photometry, spectroscopy or parallax measurements could settle current ambiguities. Given the close separations of spectral binaries like WISE~J1355$-$8258~\citep{2015AJ....150..163B}, future epochs of RV may be required to determine whether binary motion influences the kinematic analysis in this study. 

Systems analogous to WISE~J1355$-$8258 could be found in other young moving groups and star formation regions through this technique. Assuming an effective limiting magnitude of $J=19$ (appropriate for FIRE prism;~\citealt{2012PASP..124.1336S}), data to identify these systems could be obtained for distances up to 650\,pc for young M7 dwarfs, and up to 165\,pc for young L5 dwarfs. Several $1-200$\,Myr moving groups and clusters are found within 165\,pc\footnote{e.g. Taurus (1\,Myr at 140\,pc;~\citealt{2017AJ....153...46L}), Chameleon (2\,Myr at 160\,pc;~\citealt{2007ApJS..173..104L,2017MNRAS.465..760B}), TW Hydra ($7-13$\,Myr at $40-62$\,pc;~\citealt{2015MNRAS.454..593B}), Tucana Horologium ($20-40$\,Myr at $40-50$\,pc;~\citealt{2008hsf2.book..757T}), AB Doradus ($130-200$\,Myr at $20-50$\,pc;~\citealt{2015MNRAS.454..593B}).}. Identifying young spectral binaries in multiple clusters and moving groups over a range of ages would allow the exploration of planetary-mass companions as a function of age, with future confirmation and characterization with the James Webb Space Telescope.

\acknowledgments
The authors thank telescope operators for their assistance during observations.  We thank graduate student Emily Martin for engaging in helpful discussion. DBG acknowledges the Sigma Xi Foundation for grant No. G201603152061647 which provided travel funding crucial for collaboration. This publication is based on data obtained with the Magellan Telescope and the NASA Infrared Telescope Facility, operated by the University of Hawaii under Cooperative Agreement No. NNX-08AE38A with the National Aeronautics and Space Administration, Science Mission Directorate, Planetary Astronomy Program. This publication makes use of data and analysis routines from the SpeX Prism Library Analysis Toolkit (SPLAT) maintained at \url{http://www.browndwarfs.org/splat}. A.J.B. acknowledges funding support from the National Science Foundation under award No. AST-1517177. The material presented here is based upon work supported by the National Aeronautics and Space Administration under Grant No. NNX15AI75G.
\vspace{5mm}
\facilities{Magellan(FIRE).}

\software{Astropy~\citep{2013A&A...558A..33A}, Pandas~\citep{mckinney12}, SPLAT~\citep{2014ASInC..11....7B}, IDL by Harris Geospatial.}


\end{document}